# LCSH, SKOS and Linked Data

| Ed Summers | Antoine Isaac | Clay Redding | Dan Krech |
| --- | --- | --- | --- |
| Library of Congress, USA | Vrije Universiteit Netherlands | Library of Congress, USA | Library of Congress, USA |
| edsu@loc.gov | aisaac@few.vu.nl | cred@loc.gov | eikeon@eikeon.com |

## Abstract

A technique for converting Library of Congress Subject Headings MARCXML to Simple Knowledge Organization System (SKOS) RDF is described. Strengths of the SKOS vocabulary are highlighted, as well as possible points for extension, and the integration of other semantic web vocabularies such as Dublin Core. An application for making the vocabulary available as linked-data on the Web is also described.

## 1. Introduction

Since 1902 the mission of the Cataloging Distribution Service at LC has been to enable libraries around the United States, and the world, to reuse and enhance bibliographic metadata. The cataloging of library materials typically involves two broad areas of activity: descriptive cataloging and subject cataloging. Descriptive cataloging involves the maintenance of a catalog of item descriptions. Subject cataloging on the other hand involves the maintenance of controlled vocabularies like the Library of Congress Subject Headings and classification systems (Library of Congress Classification) that are used in descriptive cataloging. As Harper (2007) has illustrated, there is great potential value in making vocabularies like LCSH generally available and reference-able on the Web using semantic web technologies.

The Library of Congress makes the Library of Congress Subject Headings (LCSH) available for computer processing as MARC, and more recently as MARCXML. The conventions described in the MARC21 Format for Authority Data are used to make 265,000 LCSH records available via the MARC Distribution Service. The Simple Knowledge Organization System (SKOS) is an RDF vocabulary for making thesauri, controlled vocabularies, subject headings and folksonomies available on the Web (Miles et al., 2008). This paper describes the conversion of LCSH/MARC to SKOS in detail, as well as an approach for making LCSH available with a web application. It concludes with some ideas for future enhancements and improvements to guide those who are interested in taking the approach further.

The remainder of this paper will use LCSH/MARC to refer to Library of Congress Subject Headings represented in machine-readable format using the MARCXML format; and LCSH/SKOS will refer to LCSH represented as SKOS. A basic understanding of RDF, SKOS and LCSH is assumed for understanding the content within.

## 2. Representing LCSH as SKOS

### 2.1. Basic Model

Harper (2006) has done significant earlier work imagining LCSH/MARC as SKOS, and has provided a concrete XSLT mapping for converting MARCXML authority data to SKOS. Both SKOS and LCSH/MARC have a concept-oriented model. LCSH/MARC gathers different forms of headings (authorized/non authorized) into records that correspond to more abstract conceptual entities, and to which semantic relationships and notes are attached. Similarly SKOS vocabularies are largely made up of instances of skos:Concept, which associate a "unit of thought" with a URI. SKOS concepts have lexical labels and documentation attached to them, and can also reference other concepts using a variety of semantic relationships.

## 2.2. Concepts

Since every MARC Authority record supplied by LC contains a Library of Congress Control Number (LCCN) in the 001 MARC field, it makes a good candidate for the identification of SKOS concepts. LCCNs are designed to be persistent, and are guaranteed to be unique. SKOS requires that URIs are used to identify instances of *skos:Concept*. Semantic Web technology—as specified by RDF (Frank Manola, et al., 2004) — and Linked Data practices also encourage the use of HTTP URLs to identify resources, so that resource representations can easily be obtained (Sauermann et al., 2007). Of course LCCNs are not URLs, so the LCCN is normalized and then incorporated into a URL using the following template http://lcsh.info/{lccn}#concept.

The use of the LCCN in concept URIs marks a slight departure from the approach described by Harper (2006), where the text of the authorized heading text was used to construct a URL: e.g. http://example.org/World+Wide+Web. The authors preferred using the LCCN in concept identifiers, because headings are in constant flux, while the LCCN for a record remains relatively constant. General web practice (Berners-Lee, 1998) and more specifically recent semantic web practice (Sauermann et al., 2007) encourage the use of URIs that are persistent, or change little over time. Persistence also allows metadata descriptions that incorporate LCSH/SKOS concepts to remain unchanged, since they reference the concept via a persistent URL.

## 2.3. Lexical Labels

The MARC21 Authority format distinguishes between authorized (1XX) and non-authorized (4XX) headings. Similarly the SKOS vocabulary provides two properties, skos:prefLabel and skos:altLabel, that that allow a concept to be associated with both preferred and alternate natural language labels. In general, this allows authorized and non-authorized LCSH headings to be mapped directly to skos:prefLabel and skos:altLabel properties in a straightforward fashion.

However, a significant amount of information is also lost. The specific MARC field used to represent an authorized heading captures the type of concept: chronological (148), topical (150), geographic (151), genre/form (155). It is important for the LCSH/SKOS representation to capture the notion of different types of concepts as well (see below).

Also, a number of LCSH/MARC authorized headings are the result of combining other headings, a technique that is commonly referred to as pre-coordination. For example, a topical heading Drama can be combined with the chronological heading 17th century, which results in an LCSH/MARC record with the authorized heading Drama--17th century. In LCSH/MARC this information is represented explicitly, with original headings and subdivision 'facets'. In the LCSH/SKOS representation, headings with subdivisions are flattened into a literal, e.g. "Drama--17th century". This is an area where an extension of SKOS could be useful.

SKOS has been designed for use in a multi-lingual environment. SKOS users are encouraged to use language tags to identify the language of particular label (Isaac et al., 2008):

```
ex:animals rdf:type skos:Concept;
  skos:prefLabel "animals"@en;
  skos:prefLabel "animaux"@fr.
```

However, not all lexical labels in LCSH/SKOS are in English, e.g. Cueva de La Griega (Spain). Since this heading contains both Spanish and English it's not entirely clear what language tag to use. In addition LCSH/MARC records do not contain an indicator of what languages are used in heading fields—so it would be challenging to programmatically assign them.

### 2.4. Semantic Relationships

LCSH/MARC uses the 5XX fields to link an authorized heading to other related authorized headings. SKOS provides a rich set of semantic relationships between conceptual resources, including: skos:related, skos:broader, skos:narrower.

The semantic relationships present in LCSH/MARC are easily translated into LCSH/SKOS. The links in LCSH/MARC use the established heading as references, whereas in LCSH/SKOS conceptual resources are linked together using their URIs. This requires that the conversion process lookup URIs for a given heading when creating links. In addition LCSH/MARC lacks narrower relationships, since they are inferred from the broader relationship. When creating skos:broader links, the conversion process also creates explicit skos:narrower properties as well. Once complete conceptual resources identified with URIs are explicitly linked together in a graph structure similar to Figure 1, which represents concepts related to the concept "World Wide Web".

### 2.5. Documentation Properties

LCSH/MARC has a collection of fields that document aspects of the heading, including: general notes (667), source data (670), historical data (678), and examples (681). The SKOS vocabulary also includes documentation properties which can be used to represent LCSH/SKOS: skos:note, skos:editorialNote, skos:definition, skos:scopeNote, skos:changeNote, skos:historyNote. These properties are easily converted from LCSH/MARC to LCSH/SKOS, and require little massaging.

### 2.6. Using non-SKOS Documentation Properties

LCSH/MARC contains other features such as a relevant Library of Congress Classification Number ranges, the date that the record was created, and the date that a record was last modified. While the SKOS vocabulary itself lacks properties for capturing this information, the flexibility of RDF allows other vocabularies such as Dublin Core to be imported and mixed into SKOS descriptions: dcterms:lcc, dcterms:created, dcterms:modified. The flexibility to mix other vocabularies in to resource descriptions at will, without being restricted to a predefined schema is a powerfully attractive feature of RDF.

### 2.7. LCSH/SKOS Mapping

The general transformations above have been summarized into the following set of mappings.

| MARC Field | Feature/Function | RDF Property | Value of the Property/Comments |
|---|---|---|---|
| 010 | Control Number | rdf:about | the URI for the skos:Concept instance |
| 150 | Topical Term | skos:prefLabel | subfields: a, b, v, x, y, z |
| 151 | Geographic Term | skos:prefLabel | subfields: a, b, v, x, y, z |
| 450 | See From Tracing (Topical Term) | skos:altLabel | subfields: a, b, v, x, y, z |
| 451 | See From Tracing (Geographic Name) | skos:altLabel | subfields: a, b, v, x, y, z |
| 550 | See Also From Tracing (Topical Term) | skos:broader | only use this property when subfield w is 'g'; use value to lookup Concept URI |
| 550 | See Also From Tracing (Topical Term) | skos:related | only use this property when subfield w is not present with 'g' or 'h' in position 0 ; use value to lookup Concept URI |
| 551 | See Also From Tracing (Geographic Name) | skos:broader | only use this property when subfield w is 'g'; use value to lookup Concept URI |
| 551 | See Also From Tracing (Geographic Name) | skos:related | only use this property when subfield w is not present with 'g' or 'h' in position 0 ; use value to lookup Concept URI |
| 667 | non public general note | skos:note | subfield: a |
| 670 | Source data found | dcterms:source | subfields: a, b, u |
| 675 | Source data not found | skos:editorialNote | subfield: a |
| 678 | Biographic or historical data | skos:definition | subfields: a, b, u |
| 680 | Public general note | skos:scopeNote | subfields: a,i |
| 681 | Subject example tracing note | skos:example | subfields: a, i |
| 682 | Deleted heading information | skos:changeNote | subfields: a, i |
| 688 | Application history note | skos:historyNote | subfield: a |
| 008 | Fixed Length Data Elements | dcterms:created | positions: 0-5 |
| 005 | Date and time of last transaction | dcterms:modified | |
| 053 | LC Classification Number | dcterms:lcc | subfield: a |

## 2.8. LCSH/SKOS Illustrated

Once a given LCSH/MARC record has been converted to LCSH/SKOS an RDF graph similar to Figure 1 has been created. Note: documentation properties have been left out for display purposes.

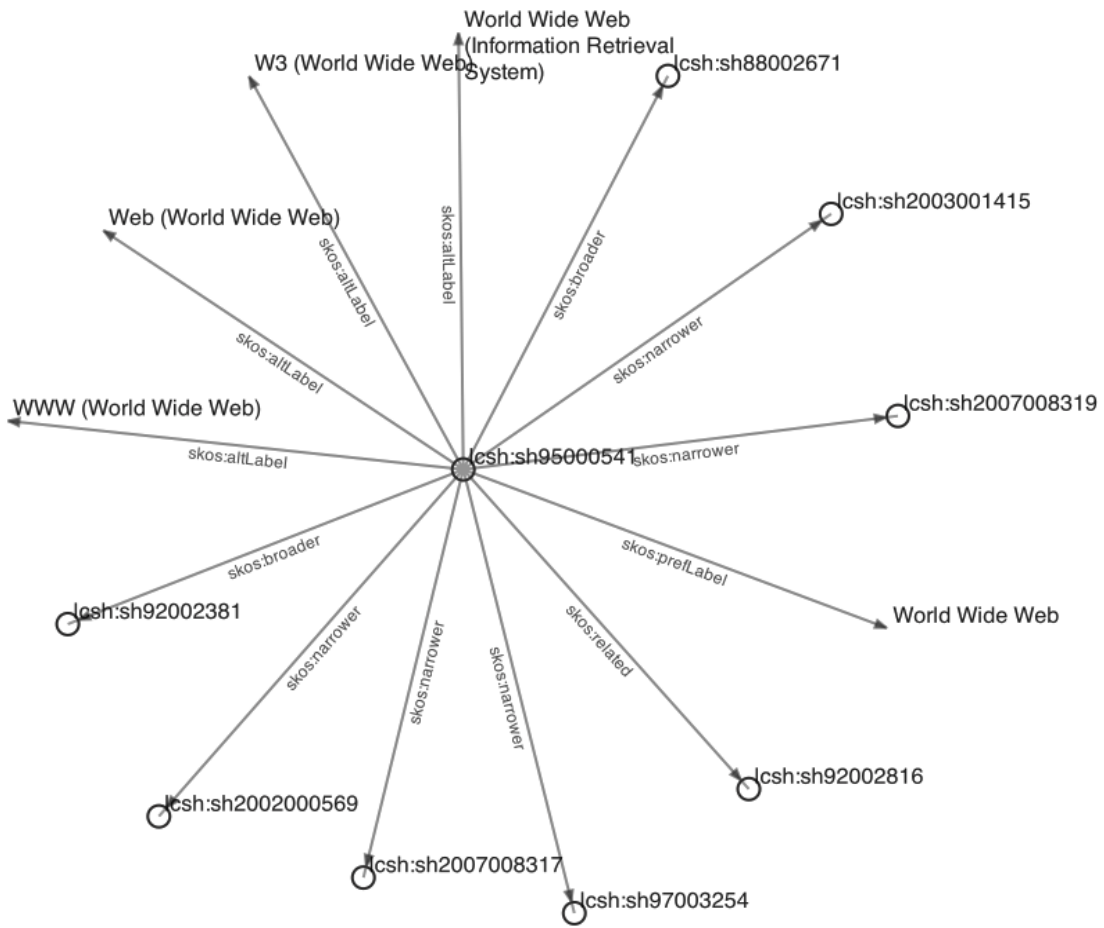

FIG. 1. SKOS Concept Graph

This example is for the concept "World Wide Web". The textual links between LCSH/MARC records are made into explicit URI links between conceptual resources, as illustrated in Figure 2

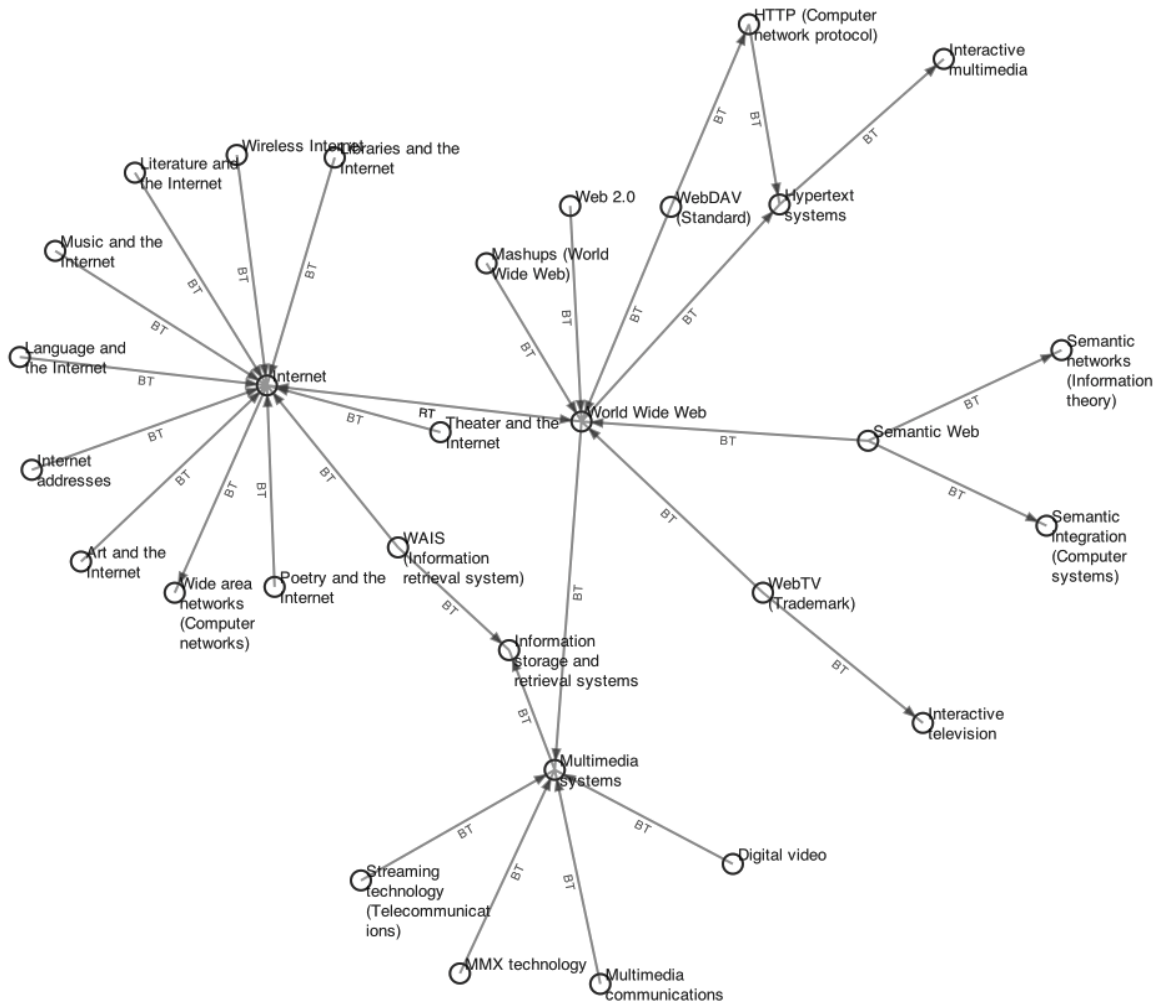

FIG. 2. Semantic Relationships Between Concepts.

## 3. Delivering LCSH/SKOS as Linked Data

### 3.1. Cool URIs for LCSH/SKOS Concepts

Implicit in the translation of LCSH/MARC to LCSH/SKOS is the minting of hundreds of thousands of URIs for conceptual resources. It is a key aspect of the semantic web and linked data (Sauermann et al., 2008) that resources are identified with resolvable HTTP URLs. The notion of "following your nose" on the World Wide Web is what allows a distributed set of machine readable descriptions to be built. The Architecture of the World Wide Web (Jacobs et al., 2004) makes a distinction between URIs for *Information Resources* (descriptions of things) and URIs for *Non-Information Resources* (the things themselves). SKOS concepts (e.g. *Mathematics*) are clearly not available on the web, so special care must be taken in minting URIs for them. Sauermann (2008) provides specific guidance on how to make resources available on the semantic web. As described in 2.2, URLs of the pattern *http://lcsh.info/{lccn}#concept* are created for each LCSH/SKOS concept. The use of hash URIs for SKOS concept simplifies the web server implementation; since the server isn't required to redirect using a *303 See Other* HTTP status code, when the URI for the concept is requested.

### 3.3. Content Negotiation

The authors chose to deliver multiple representations of LCSH/SKOS concepts on the Web using a technique called content-negotiation. When deciding what content to deliver to an HTTP client, a web server can examine the *Accept* header sent by the client, to determine the preferable representation of the resource to send (Berrueta et al, 2008). The LCSH/SKOS delivery application currently returns the following representations: rdf/xml, text/n3, application/xhtml+xml, application/json representations, using the URI patterns illustrated in Figure 3.

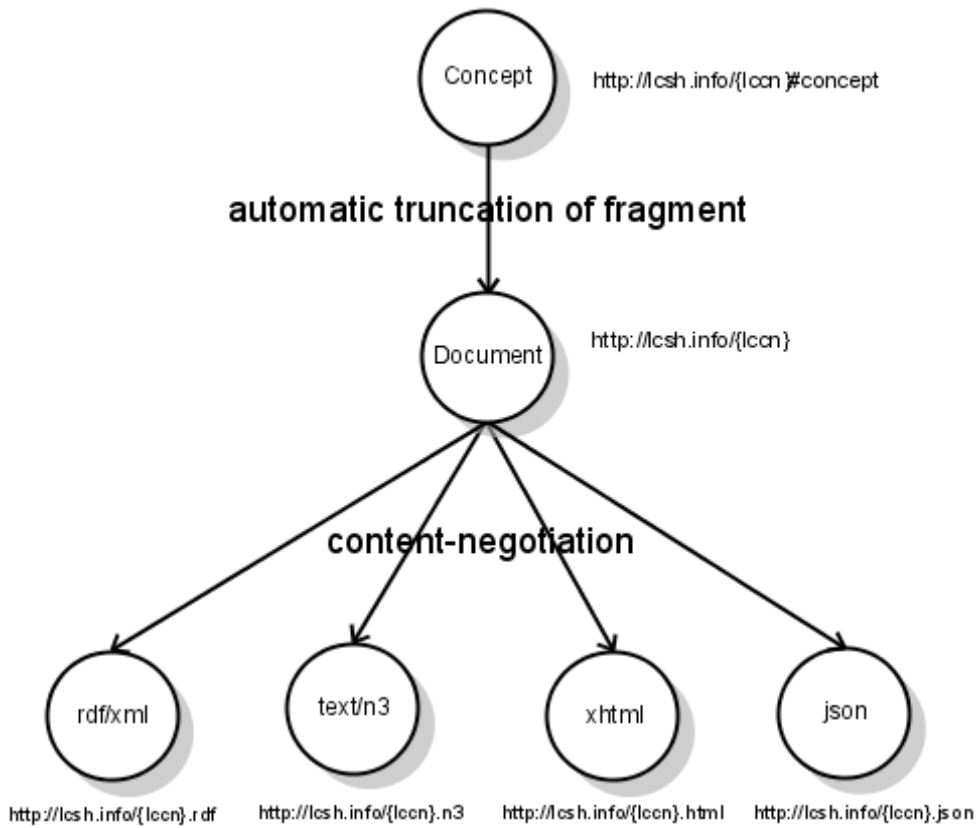

FIG. 3. URL Patterns.

The use of content-negotiation allows the LCSH/SKOS concept scheme to be browsed naturally by "following your nose" (Summers, 2008) to related concepts, simply by clicking on links in your browser (see Figure 4). It also allows semantic web and web2.0 clients to request machine-readable representations using the very same LCSH concept URIs. In addition the use of RDFa (Adida et al., 2008) allows browsers to auto-detect and extract semantic content from the human readable XHTML.

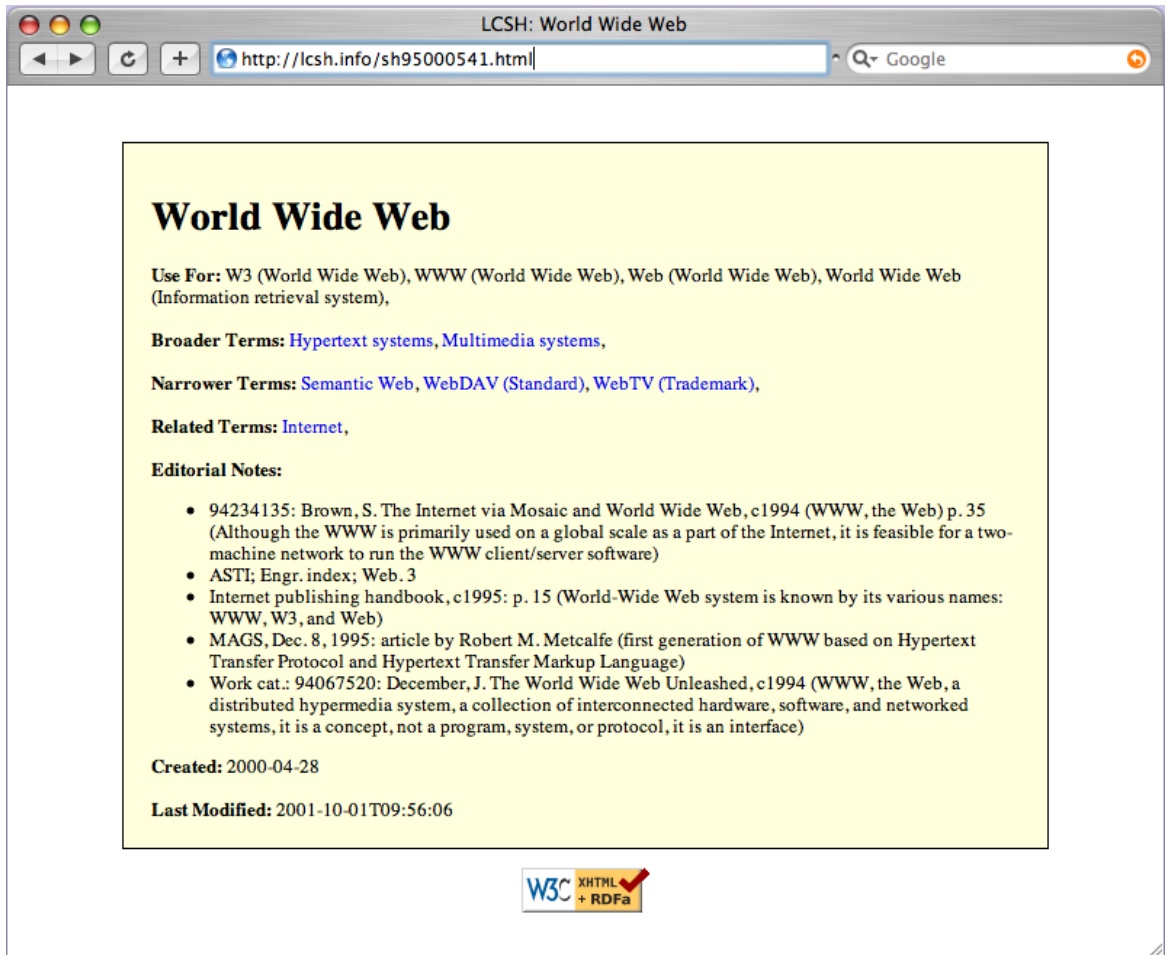

FIG 4. LCSH/SKOS Concept as RDFa XHTML.

## 4. Implementation Details

Remarkably little code (429 lines) needed to be written to perform the conversion and delivery of LCSH/SKOS. The Python programming language was used for both tasks, using a several open-source libraries:
- pymarc: for MARCXML processing (http://python.org/pypi/pymarc)
- rdflib: for RDF processing (http://python.org/pypi/rdflib)
- web.py: a lightweight web framework (http://python.org/pypi/web.py)
- webob: HTTP request/response objects, with content-negotiation support (http://python.org/pypi/WebOb)

The general approach taken in the conversion from LCSH/MARC to LCSH/SKOS differs somewhat from that taken by Harper (2006). Instead of using XSLT to transform records, the pymarc library was used, which provides an object-oriented, streaming interface to MARCXML records. In addition a relational database was not used, and instead the rdflib BerkeleyDB triple-store backend was used to store and query the 2,625,020 triples that make up the complete LCSH/SKOS dataset. The conversion process itself runs in two passes: the first to create the concepts and mint their URIs, and the second to link them together. To convert the entire dataset (377 MB) it takes roughly 2 hours, on a Intel Pentium 4 CPU 3.00GHz machine.

Readers interested in running the conversion utilities and/or the web application can check out the code using the Bazaar revision control system (http://bazaar-vcs.org) from http://inkdroid.org/bzr/lcsh.

## 5. Improvements and Future Directions

### 5.1. Extending SKOS

Since SKOS was designed as a general tool for knowledge organization systems (thesauri, classification schemes, subject heading lists, taxonomies, folksonomies) it lacks specialized features to represent some of the details found in LCSH/MARC. As discussed above in 2.3, LCSH/MARC distinguishes between several types of concepts: geographic, topical, genre/form, and chronological. However LCSH/SKOS has only one type of entity *skos:Concept* to represent all of these. As an RDF vocabulary, SKOS could easily be extended with new sub-classes of *skos:Concept: lcsh:TopicalConcept, lcsh:GeographicConcept, lcsh:GenreConcept,* and *lcsh:ChronologicalConcept*.

In addition LCSH/MARC uses pre-coordination to assemble authorized subject headings from the combination of other headings. These pre-coordinations use a variety of subfields to capture the type of facet used in a heading. Unfortunately this information is lost in SKOS since the *skos:prefLabel* property has for its range, and joins the subfields together with a '—'. Users of LCSH/SKOS will undoubtedly want to be able to identify the components of pre-coordinated concepts. Some LCSH/MARC records represent authorized subfield headings (180, 181, 182, 185), which were ignored by our initial conversion routine. It would be useful to represent these concepts using a SKOS extension. SKOS currently has an open issue (Miles, 2007) to explore how to represent coordinated concepts in SKOS, or to provide an extension pattern. Once a clear path is presented it would be useful to implement the solution in LCSH/SKOS.

### 5.2. Linking Open Data

One of the advantages of the semantic web and linked data is that traditionally isolated data sets can be integrated. Bizer (2007) provides guidance on how to link together semantic web resources using a variety of techniques. The LCSH/SKOS dataset has multiple places where links could be created to external datasets, including:

- GeoNames (http:///geonames.org) and the CIA World Fact Book (http://www4.wiwiss.fu-berlin.de/factbook/) for geographic headings.
- the RDF BookMashup (http://www4.wiwiss.fu-berlin.de/bizer/bookmashup/) for links to items that prompted a LCSH concept to be created.
- dbpedia (http://dbpedia.org)

Furthermore, there are additional vocabularies at the Library of Congress such as the Library of Congress Classification, Name Authority File, and LCCN Permalink Service which could be made available as RDF. The authors are also involved in the conversion of the RAMEAU, a controlled vocabulary that is very similar to LCSH. Once converted these vocabularies would be useful for interlinking with LCSH.

### 5.3. Server Log Analysis

Even before being announced the LCSH/SKOS web application received thousands of hits a day from web-crawling robots (Yahoo, Microsoft. Google) and semantic web applications like Zitgist and OpenLink. The server logging was adapted to also capture accept HTTP header information, in addition to referrer, user agent, IP address, concept URI. After 6 months has elapsed it will be useful to review how robots and humans are using the site: the representations that are being received, how concepts are turning up search engines like Google, Yahoo, Swoogle (http://swoogle.umbc.edu/) and Sindice (http://sindice.com).

### 5.4. Discovery with SPARQL

The LCSH/SKOS web application makes the entire dataset of 2,625,020 RDF assertions available in a single file. This dump is useful for developers who want to be able to link up their data with LCSH/SKOS concept URIs. However, given the volume of data, a SPARQL endpoint (Prud'hommeaux et al., 2008) would enable users to programmatically discover concepts without having to download and index the entire data set themselves. For example MARC bibliographic data has no notion of the LCCN for subjects that are used in descriptions. This indirection makes it impossible to determine which SKOS/LCSH concept URI to use without looking for the concept that has a given skos:prefLabel. A SPARQL service would make this sort of lookup trivial.

## 6. Conclusion

The conversion and delivery of Library of Congress Subject Headings as SKOS has been valuable on a variety of levels. The experiment highlighted the areas where SKOS and semantic web technologies excel: the identification and interlinking of resources; the reuse and mix-ability of vocabularies like SKOS and Dublin Core; the ability to extend existing vocabularies where generalized vocabularies are lacking. Hopefully the Library of Congress' mission to provide data services to the library community will provide fertile ground for testing out some of the key ideas of semantic web technologies that have been growing and maturing in the past decade.